# A 140.5µW, 99.8% Accuracy, 96.1dB SNR BioZ IC for Real-Time Cell Analysis with On-Demand Push-Pull Excitation and Continuous-Time ΔΣ T-D Readout

Jiayang Li[1], Junzhe Chen[1], Xin Zhang[2], Peiqing Li[1], Liwen Zhang[1], Navya Nayak[3], Yating Zou[4], Xuanyu Liang[3], Hubin Zhao[3], Roshan Weerasekera[1], Xiao Liu[5], Reza Haqshenas[3], Dai Jiang[3], Martin Cryan[1], Sohmyung Ha[6], Andreas Demosthenous[3], Edoardo Charbon[4], Dinesh Pamunuwa[1]

[1] University of Bristol, Bristol, United Kingdom
[2] Imperial College London, London, United Kingdom
[3] University College London, London, United Kingdom
[4] EPFL, Neuchâtel, Switzerland
[5] Fudan University, Shanghai, China
[6] New York University Abu Dhabi, Abu Dhabi, United Arab Emirates

Impedance-based phenotypic antimicrobial susceptibility testing (AST) enables label-free monitoring of bacterial responses to antibiotic exposure, supporting rapid antibiotic selection for treating bacterial infections [1]. Susceptibility is inferred from small impedance differences between treated and control samples over time and frequency [1, 2]. High SNR is required to resolve these differences, while accurate magnitude/phase readout enables reliable comparison across samples and frequencies. Moreover, parallel testing across a broad panel of antibiotic conditions further demands low power per channel (Fig. 1 (top)) [3].

Prior I/Q-based bioimpedance (BioZ) readout ICs have demonstrated SNRs of about 100dB with power consumption below 200µW, using techniques such as square-wave excitation, direct current-DAC injection, and baseline cancellation [4–6]. However, their magnitude/phase accuracy can be limited by square-wave harmonics, load-dependent current errors in direct current-DAC injection [7], and quantization and mismatch in baseline cancellation (Fig. 1 (middle left)) [8]. Prior discrete time-to-digital (T-D) readouts have demonstrated high accuracy [9–11], but generating fine timing steps using high-frequency or multiphase clocks incurs substantial clock-generation power (Fig. 1 (middle right)). For example, [11] used a 9-phase 240MHz PLL and a 10-phase DLL, which together consumed approximately 0.59mW. Combining high SNR and accurate magnitude/phase readout within a low power budget therefore remains challenging.

To address these challenges, this paper presents a BioZ IC featuring (1) an on-demand push–pull voltage driver that suppresses standing current while providing load-insensitive, low-distortion sinusoidal voltage excitation, (2) a static current comparator combining high-gain winner-take-all (WTA) current amplification with duty-cycled operation for fast, low-power current-to-time conversion, and (3) a continuous-time delta-sigma (ΔΣ) time-to-digital converter (TDC) that uses residual-time feedback to noise-shape timing quantization error, thus enabling high-resolution digitization without complex clock generation. Fabricated in a 65nm CMOS process, the proposed BioZ IC achieves a peak SNR of 96.1dB, 99.8% impedance-magnitude accuracy, 0.13° phase error and maximum power consumption of only 140.5µW.

The BioZ IC is shown in Fig. 1 (bottom). It uses a unified excitation/readout architecture with fixed-voltage excitation, eliminating a separate instrumentation amplifier (IA) and its power and noise overhead [10]. A direct digital synthesis (DDS) sinewave generator [12] uses a 6MHz clock (*CLK*) to generate $V_{sin}\pm$ at 15.6–250kHz with a programmable amplitude of 12.5–800mV$_{pp}$. An on-demand push–pull driver buffers $V_{sin}\pm$ to drive load $Z_L$ with up to 200μ$A_{pp}$ and mirrors the load current as $I_{CMP}$ for readout. In the continuous-time T-D readout, a high-speed duty-cycled static current comparator compares $I_{CMP}$ with a programmable dc reference, $I_{REF}$, to generate $CMP_{out}$. The pulse width of $CMP_{out}$ encodes the amplitude of $I_{CMP}$, while the edge midpoint encodes the phase relative to $V_{sin}^{\pm}$. An edge extractor separates the rising- and falling-edge events for two ΔΣ-TDCs, which reuse *CLK* to digitize their timings with quantization-noise shaping.

With the IA overhead eliminated by the unified excitation/readout architecture, sinusoidal load excitation dominates the remaining power budget. Even with adaptive biasing, the driver in [10] consumes 172 μA to deliver 200 μ$A_{pp}$ to the load. The proposed on-demand push–pull driver (Fig. 2 (top)) reduces standing-current overhead by supplementing a small baseline sink current, $I_{BS}$, with an on-demand auxiliary sink current, $I_{AUX}$. When the transconductance (TC) source current, $I_{SRC}$, reaches its lower limit, M7 or M8 provides $I_{AUX}$ only in the branch requiring additional sink current (Fig. 2 (bottom left)). This avoids maintaining a large standing sink current throughout the excitation cycle. Both sourcing and sinking branches operate within the TC feedback loop, which regulates $V_{\mathrm{out}}^{\pm}$ to track $V_{\mathrm{sin}}^{\pm}$ and adjusts the branch currents to compensate mismatch. DEM periodically swaps matched devices in the driver and replica output stage at $f_{\mathrm{DEM}}$, translating flicker noise out of band while averaging mismatch. The TC-to-$I_{\mathrm{CMP}}$ output stage mirrors the impedance-dependent excitation current into $I_{\mathrm{CMP}}$ through a programmable 8:N current mirror (N = 1–16), while a 5-bit I-DAC generates $I_{\mathrm{REF}}$ as the dc reference for the subsequent current comparator. Measured at 50kHz across five chips (Fig. 2 (bottom right)), the driver achieves ≤0.14% THD over a load current range of 17.5–200 μ$A_{pp}$. At 200 μ$A_{pp}$, the driver consumes 80.3μA, 53% lower than that in [10], while achieving an average current efficiency of 88%.

To resolve the crossings between $I_{CMP}$ and $I_{REF}$ with high resolution but without complex clock generation, the first continuous-time ΔΣ T-D readout front end is proposed, combining duty-cycled static comparison with residual-time noise shaping using the existing 6MHz *CLK*. The readout produces one output sample every 18 excitation cycles, with 16 cycles of comparison and T-D conversion followed by two additional TDC cycles (Fig. 3 (bottom right)). The current comparator employs a four-transistor WTA preamplifier to detect small input-current differences with a short decision delay (Fig. 3 (left)). At balance, $I_B$ is nominally split equally between $I_{WTA}^{+}$ and $I_{WTA}^{-}$. A small input difference, $\Delta I = I_{CMP} - I_{REF}$, creates unequal drain voltages at M10 and M11. Nonlinear competition between M12 and M13 amplifies this difference, steering nearly all $I_B$ to the winning output while suppressing the other toward zero. The resulting large differential current rapidly drives the cascoded stage to generate a full-swing $CMP_{out}$ transition. Polarity reversal (PR) periodically swaps the WTA input and output paths to suppress timing errors from mismatch and flicker noise while preserving the comparison polarity. The simulated propagation delay is 31–34ns for $|\Delta I|$ = 0.2–2μA, while the common delay cancels from the edge separation used for magnitude extraction. During measurements, the comparator is duty-cycled to reduce power consumption (Fig. 3 (bottom right)). Full-window comparison in cycle 1 locates both crossings. In cycles 2–16, $EN_{CMP}$ controls the switches to enable comparison only within narrow windows around subsequent crossings. The measured average comparator current is 3.1μA.

After comparison, an edge extractor converts $CMP_{out}$ into rising- and falling-edge events ($P_{in}^{+}$ and $P_{in}^{-}$) while suppressing noise-induced retriggering. Two identical ΔΣ-TDCs (Fig. 3 (top right)) digitize their timings into $D_{out}^{+}$ and $D_{out}^{-}$. During cycles 1–16, Phase Quantizer 1 (PQ1) captures a *CLK*-driven counter value as the coarse code $D_1$ and generates a residual pulse, $P_{out}$, whose width represents the interval from $P_{in}$ to the next *CLK* edge.

The $D_1$ codes are summed into $D_{sum}$, while the Residual-Time Accumulator (RTA) accumulates the $P_{out}$ pulse widths. In cycle 17, the RTA performs an internal time transfer. In cycle 18, it outputs the accumulated residue as $P_{acc}$, whose width is then quantized by PQ2 into $D_{res}$, with the remaining time residue retained as $P_{res}$. The output is $D_{out} = D_{sum} - D_{res}$. $P_{res}$ is fed back and subtracted from the next frame's accumulated residue before PQ2 quantization, realizing first-order quantization-noise shaping. A cycle-by-cycle numerical example is shown in Fig. 3 (middle right).

The PQ (Fig. 4 (bottom left)) uses a counter clocked by *CLK* and reset by *REF* (an excitation-synchronous reference signal generated by the DDS). Upon each input pulse $P_x$, the PQ generates a local *SAMP* pulse to latch the counter value as $D_x$ and also produces the corresponding residual pulse $P_y$. In each of the RTA's two time registers (TR1 and TR2), $C_{TR}$ is reset to $V_{DD}$ and discharged by $I_{TR}$ during $T_{in}$ pulses, storing the accumulated interval as a voltage drop at $V_{TR}$ (Fig. 4 (top)) [13]. For an interval $T_{acc}$ stored in TR1, *XFER* resumes discharge to the inverter threshold, producing $T_1 = T_{TR} - T_{acc}$, where $T_{TR} = C_{TR}\Delta V_{TR}/I_{TR}$. TR2 stores $T_1$ and is subsequently read out by *ACC*, producing $T_2 = T_{TR} - T_1 = T_{acc}$ for matched TRs (Fig. 4 (middle)). These two successive subtractions recover the accumulated interval as $P_{acc}$. Inter-TR mismatch introduces a timing offset whose sign reverses with the transfer order. Therefore, $\varphi_{PR}$ swaps the TR roles after each $P_{res}$, canceling this offset over each PR pair and chopping TR flicker noise out of band, while the associated routing preserves the subtractive $P_{res}$ feedback. The measured time-noise density (Fig. 4 (bottom right)) exhibits an approximately +20dB/dec high-frequency slope, consistent with first-order noise shaping; the discrete tones arise from periodic PR/DEM modulation. The measured effective timing resolution is 53.6ps at 4kΩ load (0.1–4Hz).

Measurements across five chips, which were fabricated in a 65nm CMOS process, show a peak SNR of 96.1dB at 4kΩ, with noise integrated over 0.1–4Hz (Fig. 5 (top left)). Fig. 5 (top right) shows magnitude-noise PSDs normalized by the squared mean impedance magnitude for 56Ω and 4kΩ loads at 200μ$A_{pp}$ and a 20kΩ load at 40μ$A_{pp}$. To evaluate measurement errors, the IC was tested with resistive loads spanning 56Ω–20kΩ using five operating settings. Gain mismatch was calibrated beforehand for each setting. The IC achieves a mean magnitude error of 0.24% relative to a Keithley DMM6500 (Fig. 5 (bottom left)). To validate readout accuracy with biological samples, the IC was tested with ARPE-19 epithelial cells at 60%, 70%, and 95% nominal confluence over 15.6–250kHz, after calibration with a fixed RC reference. The mean magnitude and phase errors are 0.59% and 0.13°, respectively, relative to a Zurich Instruments MFIA impedance analyzer (Fig. 5 (bottom right)). At 1.2V supply, the complete IC, including the DDS, driver, and readout, consumes a maximum of 140.5μW. Fig. 6 compares the proposed IC with prior BioZ readout ICs. Among the compared designs, this work achieves the lowest reported maximum power consumption while delivering high magnitude/phase accuracy and competitive SNR. Fig. 7 shows the die micrograph and power breakdown.

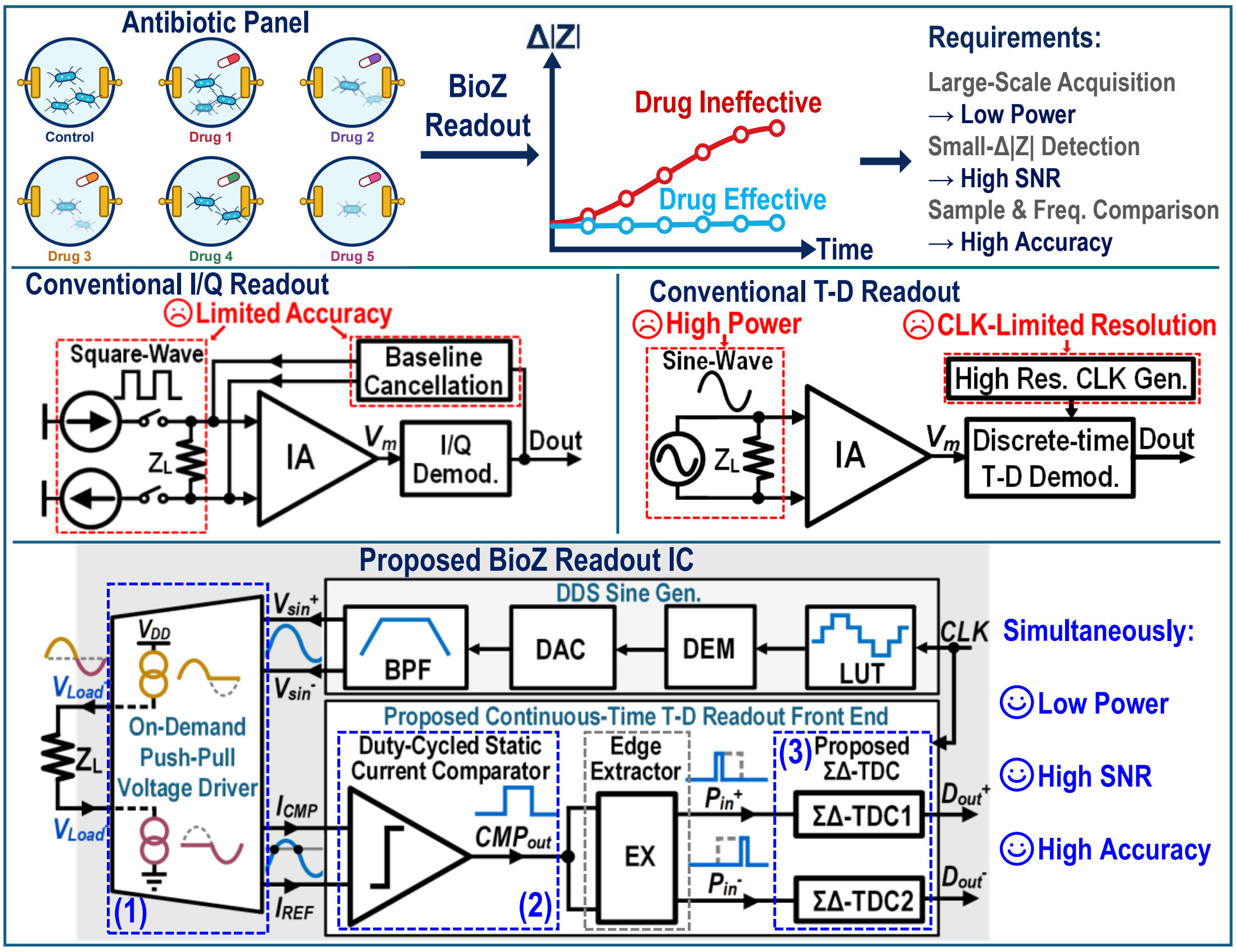

**Fig. 1: BioZ-based antimicrobial-susceptibility testing, limitations of prior I/Q and T-D readouts, and block diagram of the proposed continuous-time T-D BioZ IC.**

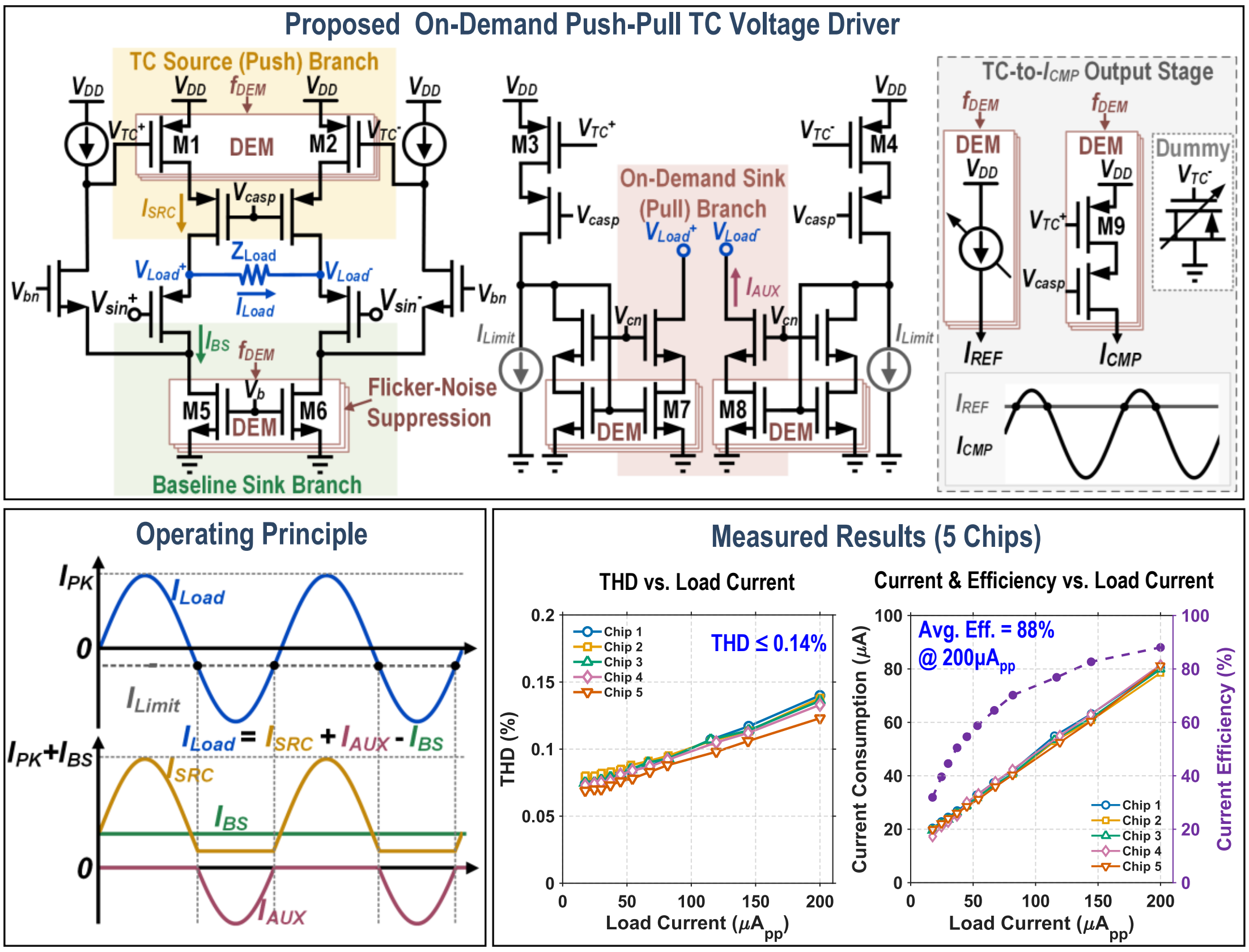

**Fig. 2: Proposed on-demand push–pull TC voltage driver, operating principle, and measured THD, current consumption, and current efficiency across 5 chips.**

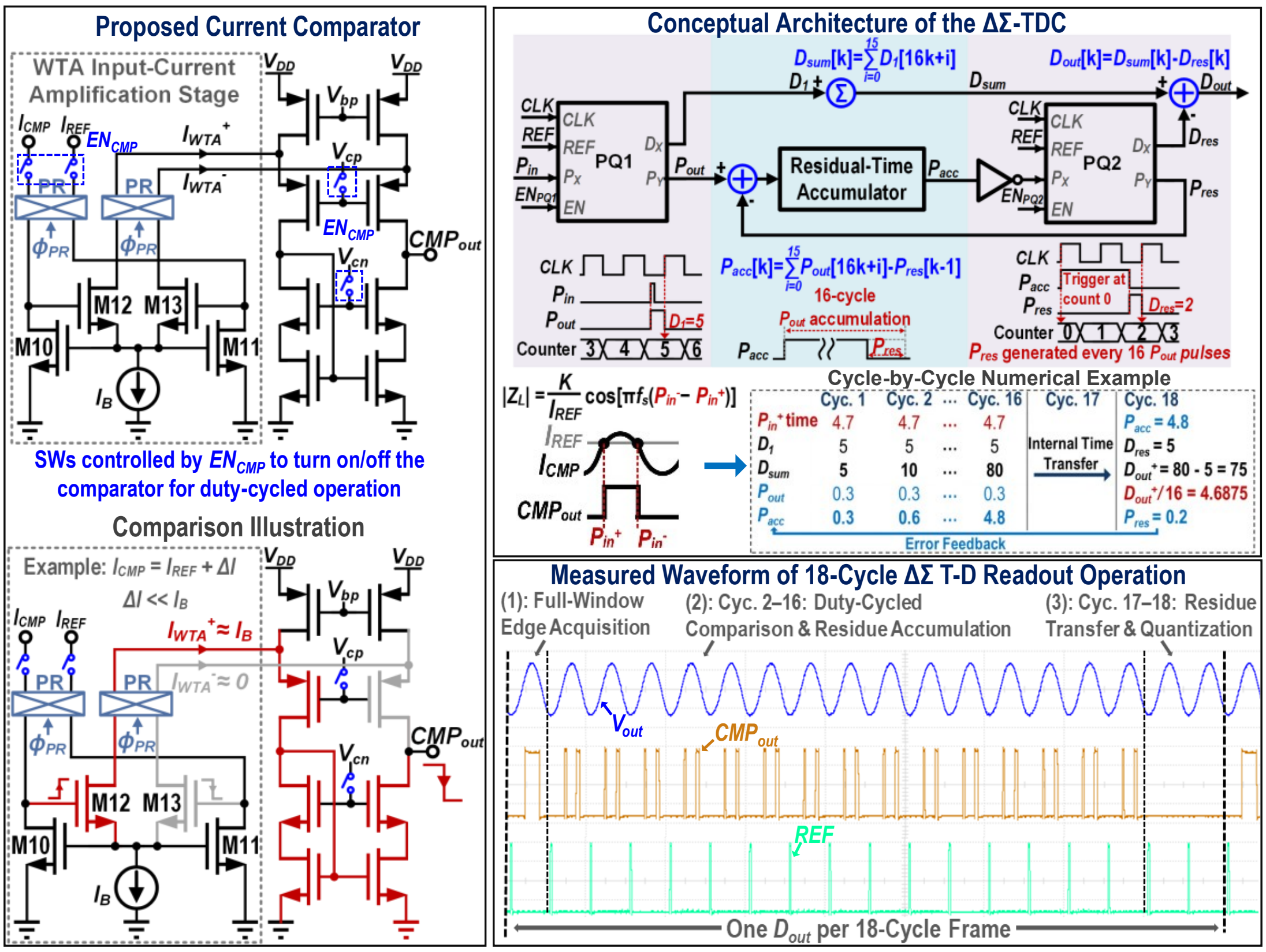

Fig. 3. Proposed duty-cycled static current comparator, conceptual ΣΔ-TDC architecture, and measured 18-cycle T-D readout operation.

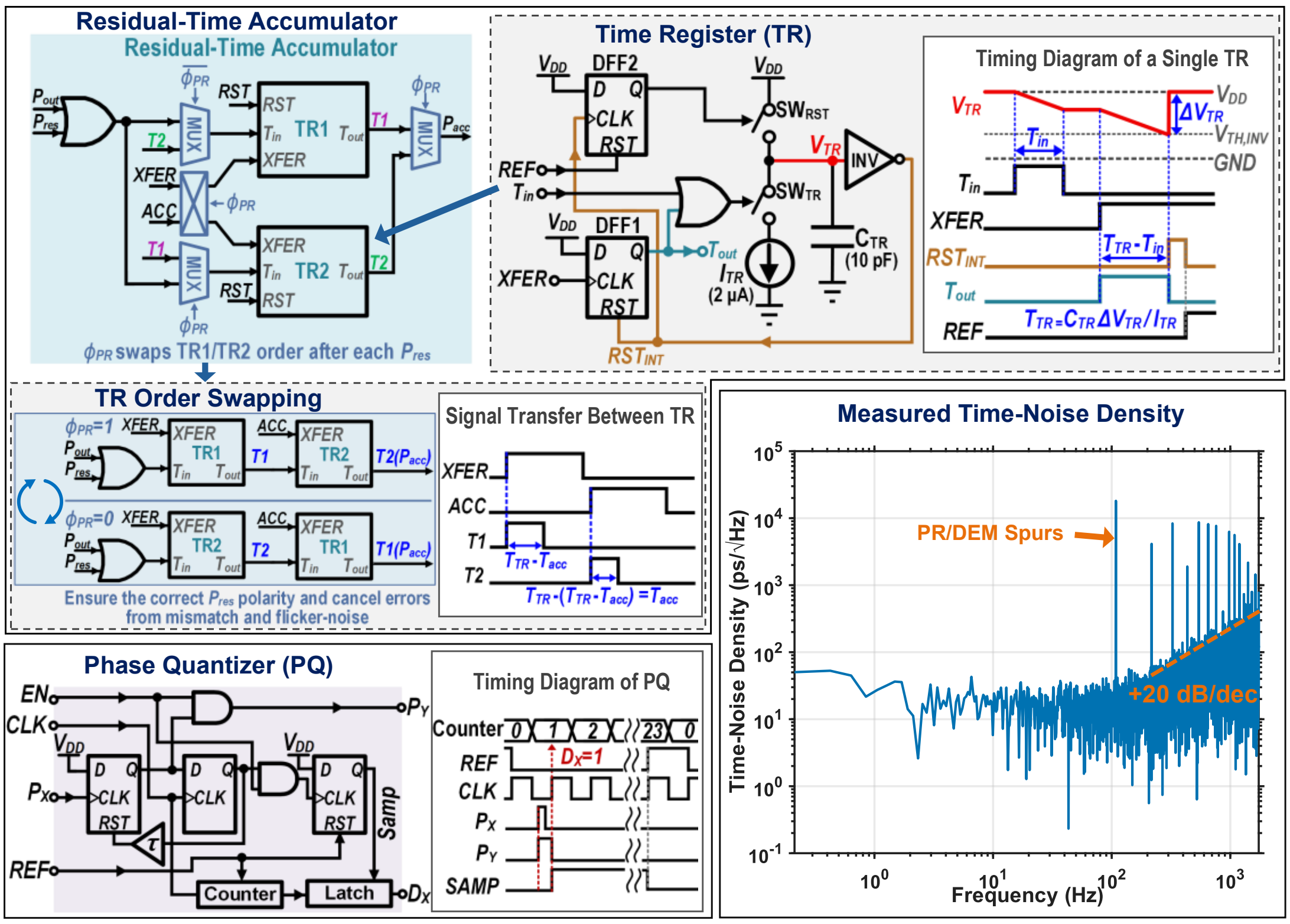

Fig. 4. Proposed residual-time accumulator, time register, TR order swapping, phase quantizer, and measured time-noise density.

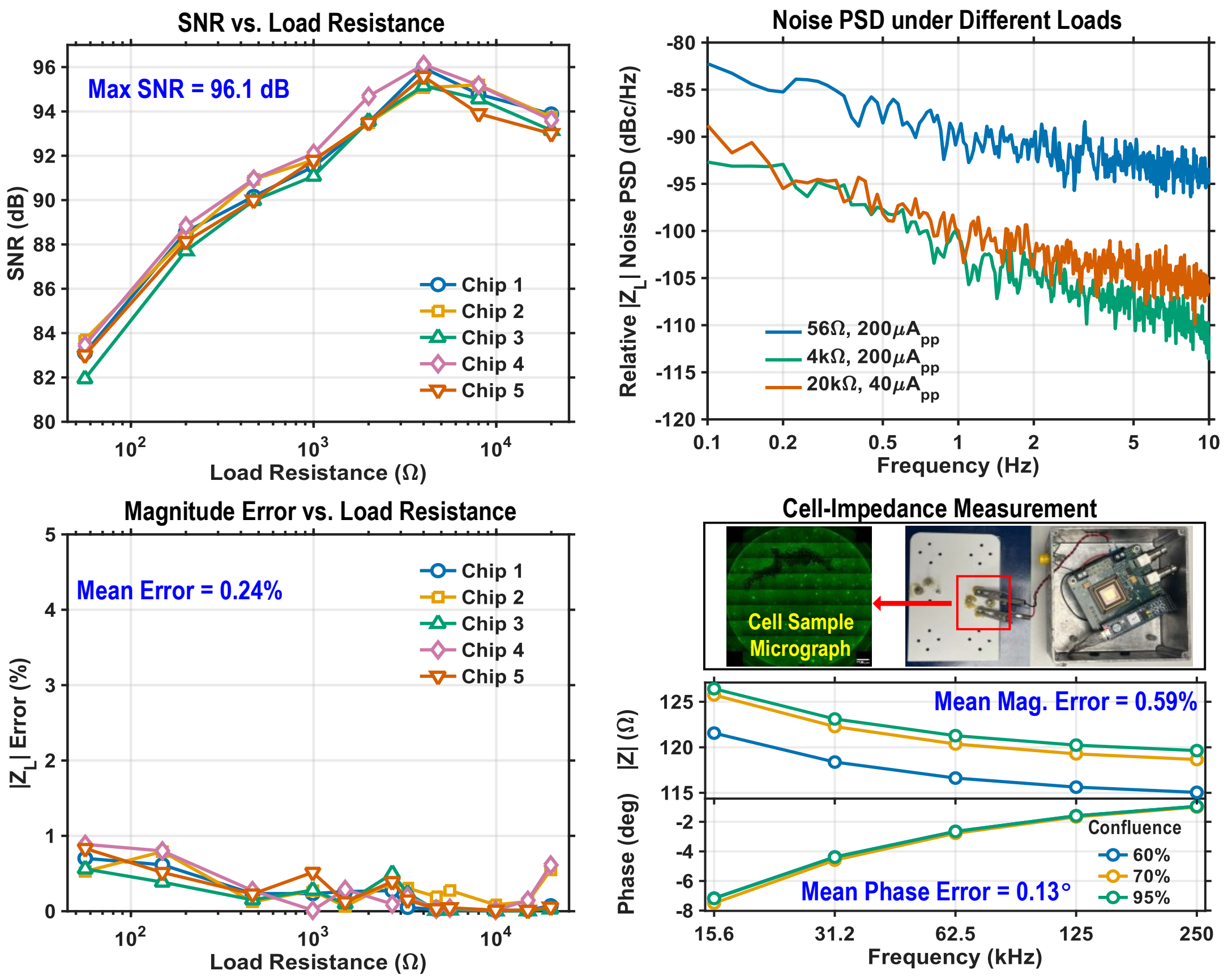

**Fig. 5. Measured BioZ SNR, magnitude error, impedance-noise spectra, and cell-impedance results.**

| Parameters | ISSCC'22[5] | ISSCC'24[6] | JSSC'25[14] | TBCAS'25[15] | ISSCC'25[11] | This Work |
|---|---|---|---|---|---|---|
| Technology (nm) | 180 | 180 | 180 | 180 | 65 | 65 |
| Supply (V) | 1.2 | 1.2/0.9 | 1.2/1.0 | 1.8 | 1.2 | 1.2 |
| Readout Method | I/Q | I/Q | I/Q | I/Q | Discrete T-D | CT ΔΣ T-D |
| Excitation Signal | Square | Pseudo-Sine | Pseudo-sine | Pseudo-sine | Sine | Sine |
| Current Magnitude ($\mu A_{pp}$) | 10–200 | 6–200 | 12.5–200 | 10–100 | 15-500 | Up to 200 |
| Frequency Range (kHz) | 1–1024 | 1–200 | 1–500 | 1–100 | 10–30000 | 15.6–250 |
| Load Range (Ω) | 20–4k | ≤37k | 20–4k | ≤36k | 10–1k | 56–20k |
| Max Power Consumption (μW) | 144.2[a] | 160.7[a] | 255.9 | 175.5 | 3180 | 140.5 |
| Max SNR (dB) | 106 | 101.5 | 97.3 | 91.1 | 92.7 | 96.1[b] |
| Phase Error (°) | N/A | N/A | N/A | ≤0.48 | 0.57 | 0.13[c] |
| Mag. Error (%) | N/A | N/A | N/A | ≤3.3 | 0.39 | 0.24[d] / 0.59[c] |

[a]Calculated from the paper.

[b]SNR = $20\log_{10}$ (mean impedance magnitude / rms impedance noise), with noise integrated over 0.1–4 Hz.

[c]Mean errors measured using cell samples.

[d]Mean errors measured using resistive loads.

**Fig. 6. Performance comparison with prior BioZ readout ICs.**

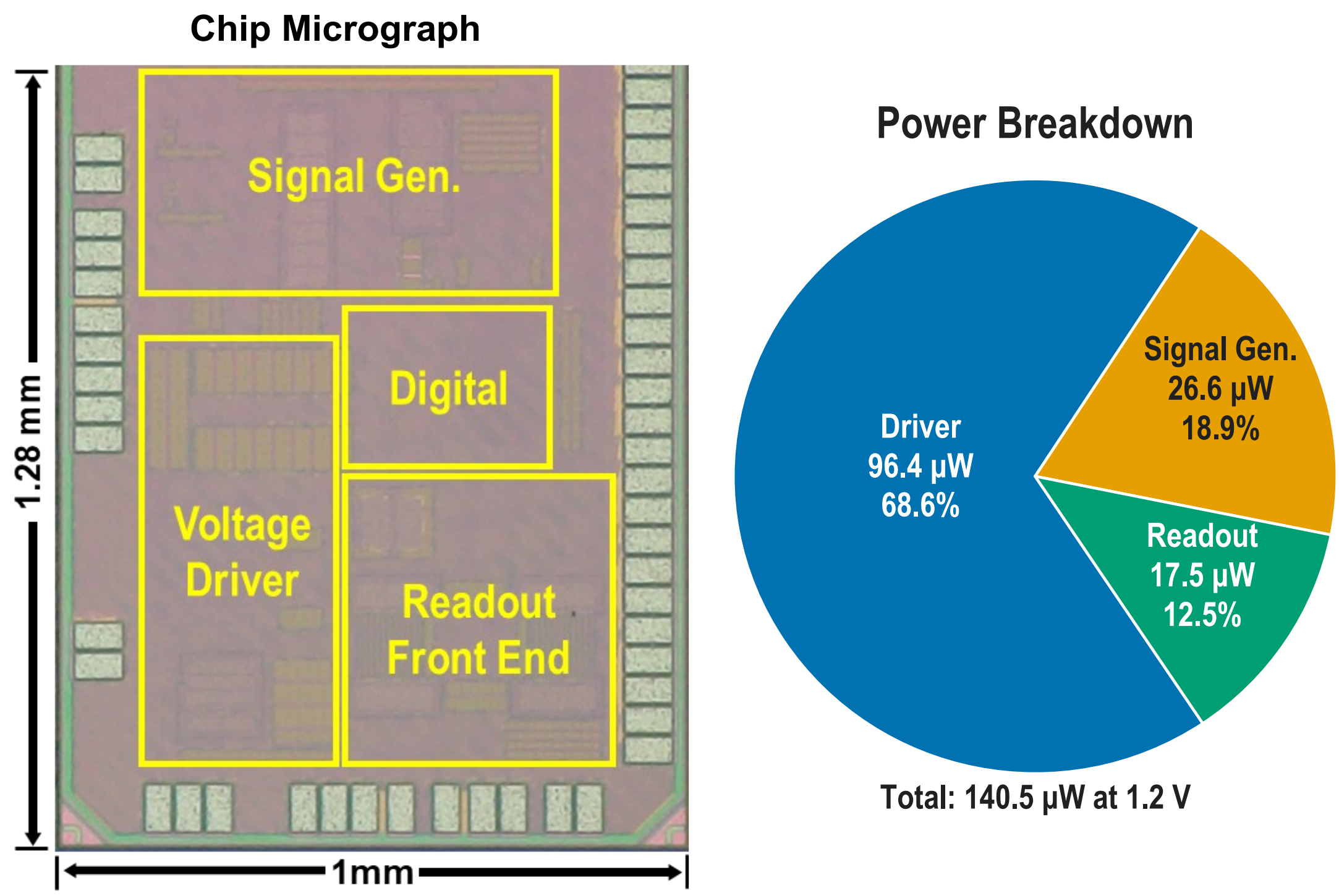

**Fig. 7. Die micrograph and power breakdown of the proposed BioZ IC.**